# Quasi-average predictions and regression to the trend: an application the M6 financial forecasting competition

Jose M. G. Vilar[1,2,*]

[1] Biofisika Institute (CSIC, UPV/EHU), University of the Basque Country (UPV/EHU), P.O. Box 644, 48080 Bilbao, Spain

[2] IKERBASQUE, Basque Foundation for Science, 48011 Bilbao, Spain

[*]Correspondence to: j.vilar@ikerbasque.org

## Abstract

The efficient market hypothesis considers all available information already reflected in asset prices and limits the possibility of consistently achieving above-average returns by trading on publicly available data. We analyzed low dispersion prediction methods and their application to the M6 financial forecasting competition. Predictive averages and regression to the trend offer slight but potentially consistent advantages over the reference indexes. We put these results in the context of high variability approaches, which, if not accompanied by high information content, are bound to underperform the benchmark index as they are prone to overfit the past. In general, predicting the expected values under high uncertainty conditions, such as those assumed by the efficient market hypothesis, is more effective on average than trying to predict actual values.

## 1. Introduction

The efficient market hypothesis contemplates financial markets as informationally efficient, with all available information already reflected in asset prices (Fama, 1970; Malkiel, 2019), implying that it is impossible to consistently achieve above-average returns by trading on publicly available information, as any new information is rapidly incorporated into stock prices. Consequently, this hypothesis challenges the notion of consistently outperforming the market and renders most forms of fundamental and technical analysis eventually futile (Samuelson, 1974).

Indeed, over the past 10 and 20 years, less than 7% and 10%, respectively, of U.S. actively managed equity funds have outperformed their benchmarks (Edwards, Ganti, Lazzara, Nelesen, & Di Gioia, 2023). Since the conclusion of 2022, the most recent instance when the average active U.S. stock fund outperformed the S&P 500 stock index for an entire calendar year was in 2009.

This type of underperformance was also highlighted in a study of all 2,132 U.S. actively managed funds, except for those concentrating on niche market segments or employing leverage, from June 2017 to June 2022 (Lazzara, Stoddart, & Di Gioia, 2023), which found that none of the funds maintained a top-quarter performance ranking each year over the entire five-year period. Relaxing the criterium to a top-half performance in each of the five years under examination resulted in less than 1% of all funds doing so. Remarkably, this value is lower than the 6.5% expected persistence under random distribution (the probability of the top half remaining on the top for the next four years is $0.5^4 = 0.065$). This effect indicates that high performers tend to perform worse than random later on, which would be consistent with a regression to the mean effect (Stigler, 1997).

Therefore, forecasting methods, along with the efficacy of various trading strategies, are severely challenged by efficient markets, as prices are already a reflection of all available information. However, there can potentially be some signals from the inherent asset price dynamics and economic indicators evolution that could be used. These challenges and potential signals further question whether indeed actively managed portfolios can consistently outperform passive index funds in the long run.

The M6 Financial Forecasting Competition is the 6th installment of the Makridakis competitions (Makridakis, Fry, Petropoulos, & Spiliotis, 2022; Makridakis, et al., 2023). Its focus on forecasts of stock returns and risk lies at the center of the efficient market hypothesis. It asked participants to forecast the relative performance of multiple assets and their investment positions from a universe of 50 S&P500 stocks and 50 international Exchange-Traded Funds (ETFs), encompassing diverse categories and geographical areas. The relative performance track consisted of forecasting the rank of each asset returns. The investment position track consisted of developing a portfolio, with potentially long and short positions, to obtain maximum returns adjusted by the risk. The M6 petition spanned 12 submission periods of four weeks each.

Here, we present an approach that outperformed the benchmark indices in both the relative performance and the investment positions and that ranked 5th overall in their combined score in the M6 competition. The approach is based on the idea that, in high variability contexts, it is often more effective to predict values close to the expected averages of categories and trends than to try to make precise predictions. The approach considered only data within the M6 competition universe, namely, the historical values of the 100 assets.

Explicitly, the focus is on methods that combine average values. For forecasting performance, we considered the average performance distribution of stocks and ETFs separately and combined it with the average performance distribution over time for each specific asset. The motivation stems from the property of the average value as the quantity that minimizes the mean squared error of a set of random variables. Substantial departures from the average value would not be beneficial unless they are highly informative. For the investment decisions, the strategy consisted of selecting the stocks that performed better in the long term but not too well in the short term. Discarding short-term overachievers is intended to remove recent fluctuations that would potentially regress to the trend. To further suppress the variability resulting from the reference index, we shorted ETFs. The motivation for this type of shorting is that, although it would reduce the returns, it would also reduce the volatility, which leads to a higher information ratio for historical data.

## 2. Forecasting performance

The forecasting performance was measured by the Ranked Probability Score ($RPS$), as detailed in (Makridakis, et al., 2023). In short, the $RPS$ for asset $i$ in the period $T$ is calculated as

$$RPS_{i,T} = \frac{1}{5}\sum_{j=1}^{5}\left(\sum_{k=1}^{j} q_{i,T,k} - \sum_{k=1}^{j} f_{i,T,k}\right)^2 . \tag{1}$$

Here, the ranking of the return of asset $i$ at time $T$ is described by a vector $q_{i,T}$ with components $q_{i,T,k}$ for $k \in 1, \dots, 5$, which are set equal to one if the asset is ranked in the quintile $k$ and zero otherwise. The forecasted probabilities of each rank for a particular asset are denoted by $f_{i,T,k}$. The overall $RPS$ for multiple submission points from $T_1$ to $T_2$ is

$$RPS_{T_1:T_2} = \frac{1}{100(T_2 - T_1 + 1)} \sum_{T=T_1}^{T_2} \sum_{i=1}^{100} RPS_{i,T} . \tag{2}$$

The $RPS$ is non-negative in general and reaches the zero value for a perfect score. The method that relies on the least information, used as a forecasting benchmark, considers the probabilities of each quintile being realized equal to 0.2 for all assets (i.e., each asset is equally likely to have a performance in any of the quintiles). It leads to an $RPS$ equal to 0.16. It is used as a benchmark because it is consistent with the efficient market hypothesis and is difficult to improve by a large margin. Other naïve approaches perform substantially worse. For instance, predicting only the 1st, 2nd, 3rd, 4th, or 5th quintile with probability one for all the assets would lead to a $RPS$ of 0.4, 0.28, 0.24, 0.28, and 0.4 respectively, all of which are higher than the benchmark. Randomly predicting any quintile with probability 1 would lead on average to an $RPS$ of 0.32. These naïve approaches also indicate that predicting further away from the expected averages worsens the forecast performances in high variability scenarios.

The way variability affects nonlinear systems is complex and often paradoxical (Vilar & Rubi, 2001). To gain insights into the effects of high variability, it is useful to investigate the conditions for optimality of the average performance. Consider the average $RPS$ expressed as

$$\langle RPS_{i,T}\rangle = \frac{1}{5}\sum_{j=1}^{5}\left(\langle Q_{i,T,j}\rangle - F_{j,T,k}\right)^2$$

(3)

with $Q_{i,T,j} = \sum_{k=1}^{j} q_{i,T,k}$ and $F_{i,T,j} = \sum_{k=1}^{j} f_{i,T,k}$. The condition of extremum is

$$\frac{\partial}{\partial F_{j,T,k}}\langle RPS_{i,T}\rangle = -\frac{2}{5}\left(\langle Q_{i,T,j}\rangle - F_{j,T,k}\right) = 0,$$

(4)

which implies $F_{j,T,k} = \langle Q_{i,T,j}\rangle$ and equivalently $f_{i,T,k} = \langle q_{i,T,k}\rangle$. The extremum is a minimum because $\frac{\partial^2}{\partial F_{j,T,k}^2} RPS_{i,T} = \frac{2}{5}$ is positive.

The estimation of $\langle q_{i,T,k}\rangle$, however, is challenging because of changing conditions. In general, low variability approaches are expected to perform better on average because the nonlinearity introduced by the evaluation metric results in a convex function (Vilar & Rubi, 2018). The requirements can be clarified by considering the forecast variability as $e_{j,T,k} = F_{j,T,k} - \langle F_{j,T,k}\rangle$, which is zero on average by definition. Using $F_{j,T,k} = \langle F_{j,T,k}\rangle + e_{j,T,k}$ in the expression for $RPS_{i,T}$ leads to

$$\langle RPS_{i,T}\rangle = \frac{1}{5}\sum_{j=1}^{5}\left(\langle Q_{i,T,j}\rangle - \langle F_{j,T,k}\rangle\right)^2 - \frac{2}{5}\sum_{j=1}^{5}\langle Q_{i,T,j} e_{j,T,k}\rangle + \frac{1}{5}\sum_{j=1}^{5}\langle e_{j,T,k}^2\rangle,$$

(5)

which indicates that variability in the forecast needs to be highly correlated with the rank of the realized returns to increase the performance (lover the expected $RPS$). Explicitly, the term $\sum_{j=1}^{5}\langle Q_{i,T,j} e_{j,T,k}\rangle$ has to be larger than $\frac{1}{2}\sum_{j=1}^{5}\langle e_{j,T,k}^2\rangle$.

Based on these considerations, the key elements are to use $f_{i,T,k} = \langle q_{i,T,k}\rangle$ and to estimate $\langle q_{i,T,k}\rangle$ with as little variability as possible. A potential approach is to make asset classes and to estimate the expected value as the observed average of the class over a short period of time. In general, ETFs have lower volatility than stocks, which would make the 1st and 5th return quintiles to be more present in stocks than in ETFs. Indeed, this behavior is observed within the M6 competition universe, with the salient exception of the VXX ETF (Fig. 1).

Splitting the average of $q_{i,T,k}$ over one period of the M6 competition (four weeks) into $i \in stocks$ and $i \in ETFs$, just two types of assets, leads to a fluctuating $RPS$ that is generally lower than the benchmark (Fig. 2). This performance was assessed as the moving average of the $RPS$ for 3 periods, a quarter of the M6 competition duration, and 12 periods (48 weeks), the whole duration of the M6 competition.

Another potential approach to estimate $\langle q_{i,T,k}\rangle$ is to consider a temporal average of each asset. The tradeoff is that short averaging intervals would lead to higher variability and longer ones would include outdated conditions. The approach considered averages over the last 5, 10, and 400 periods, weighted each by 0.2, 0.2, and 0.6, respectively. The results are generally better than the benchmark as well (Fig. 2).

The results can be improved further by estimating $\langle q_{i,T,k} \rangle$ as the average of both the asset-type and temporal averages (Fig. 2). This approach, with slightly different parameters in each submission, ranked first in the first quarter of the M6 competition and eighth globally out of 163 teams, with an $RPS$ equal to 0.1573, which was very close to best performer value of 0.1565.

## 3. Investment decisions

The investment decisions consisted of selecting a portfolio to obtain maximum returns adjusted by the risk. The evaluation was performed as detailed in (Makridakis, et al., 2023). In short, the performance from day $t_1$ to $t_2$ is measured by

$$IR_{t_1:t_2} = \frac{ret_{t_1:t_2}}{sdp_{t_1:t_2}}, \tag{6}$$

where $IR$ is referred to as the information ratio, $ret$ represents continuously compounded returns, and $sdp$ is the standard deviation of the daily returns. $IR$ can be viewed as a variant of the conventional Information Ratio, but with the benchmark return set equal to 0. It is also a variant of the Sharpe Ratio with the risk-free rate set to 0.

The initial strategy consisted of selecting the stocks that performed better than half of the assets over 120 trading days. Additionally, stocks that performed better than 85% of the assets over 40 trading days were discarded. Namely, the strategy selects long-term performance but considers high short-term performance as a detrimental effect that would potentially be corrected to maintain the trend. This correction is analogous to the regression to the mean phenomenon, which indicates that if a random variable has an extreme value, the next value is likely to approach its mean (Stigler, 1997). The reversion can occur for various reasons, including overreaction or irrational market behavior. Investors who react to false information or make decisions too late may experience these mean-reverting movements as markets correct themselves.

In mathematical terms, to select $w_i$, the approach considers the nonnormalized weights

$$v_{i,t} = \mathbf{I}\left\{ i \in stocks \ \wedge \ \mathrm{rank}\left(\frac{S_{i,t-1}}{S_{i,t-121}}\right) > 50 \ \wedge \ \mathrm{rank}\left(\frac{S_{i,t-1}}{S_{i,t-41}}\right) \leq 85 \right\}, \tag{7}$$

where $\mathbf{I}$ is the indicator function, which is 1 if its argument is true and 0 otherwise. Normalization leads to the portfolio weights

$$w_{i,t}^{\mathrm{RtT}} = \frac{v_{i,t}}{\sum_i v_{i,t}}. \tag{8}$$

Here, the superscript RtT refers to the regression to the trend approach.

In terms of returns, the regression to the trend approach performs generally better than the benchmark, especially over long investment horizons (Fig. 3). In terms of the $IR$, the approach also performs better than the benchmark, although the improvement is not as marked (Fig. 4).

A subsequent strategy to improve the $IR$, not considered initially in the competition, was to compensate fluctuations with short positions in ETFs. Given the nonnormalized weights for ETFs $u_i = \mathbf{I}\{i \in ETFs\}$, the overall weights of the portfolio were selected as

$$w_{i,t}^{\text{Compensated}} = \frac{2}{3} \cdot \frac{v_{i,t}}{\sum_i v_{i,t}} - \frac{1}{3} \cdot \frac{u_i}{\sum_i u_i}. \tag{9}$$

Namely, two thirds of the portfolio were long positions in selected stocks and one third consisted of short positions in ETFs. This approach, referred to as compensated, provided returns close to the benchmark (Fig. 3) but with $IRs$ higher than those of both the benchmark and the regression to the trend approach (Fig. 4).

The choice of shorting just ETFs, rather than stocks or stocks and ETFs, was based on a slightly better performance over 48 periods (Fig. 5). In general, the performance of each shorting type over 12 or 3 periods was essentially the same, with only marginal fluctuating differences.

The regression to the trend approach ranked 27$^{\text{th}}$ the first quarter of the competition and switching afterwards to the compensated approach ranked 36$^{\text{th}}$ globally out of 163 teams, with a global $IR$ of 1.301, which was higher than the value of 0.4535 obtained by the benchmark. The combined performance of the approaches presented above ranked 3$^{\text{rd}}$ in the first quarter and 5$^{\text{th}}$ globally out of 163 teams.

## 4. Discussion

The results of the M6 competition have shown that, over its 48-week overall duration, the majority of the teams underperformed the benchmarks: only 23.3% provided more precise forecasts, just 28.8% developed better portfolios, and merely 6.7% achieved both of them (higher $IR$ and $RPS$ scores) (Makridakis, et al., 2023). These results are analogous to the general tendency of most actively managed equity funds to underperform their benchmarks. The underperformance of actively managed funds has often been attributed to high management costs, limitations in high-volume trading, fees, and risk-managed portfolios, among others (Del Guercio & Reuter, 2014). These limitations were not present in the M6 competition and yet the results were aligned with those observed in actual trading settings.

Here, we have shown that predicting the expected values under high uncertainty conditions, such as those assumed by the efficient market hypothesis, is more effective on average than trying to predict actual values. In general, there is a tradeoff between fine- and coarse-grained detail. Finer detail would produce more accurate results but also increase the variability of the average estimates. The information used in the approaches was restricted to the 100 assets of the M6 competition universe and the methodology was relatively straightforward (moving averages, asset-class averages, and ranking). Despite these inherent constraints, the approach outperformed both benchmarks and ranked fifth out of 163 teams in the global combined performance.

Whether investment decisions can benefit from forecasting the relative performance of multiple assets is an outstanding question. The results presented here show that forecasting performance is largely non-actionable because improvements over the naïve benchmark arise from symmetric quintile distributions, e.g., from the 1$^{\text{st}}$ and 5$^{\text{th}}$ return quintiles to be more prominent in stocks than in ETFs but with only minor biases towards any of these quintiles.

Potential avenues to improve forecasting performance would need more sophisticated methods than averaging to estimate the expected values, such as Bayesian approaches to combine the joint average performance distribution across time for each specific asset with that of their class, whether Stock or ETF. In this regard, rather than considering two classes of assets, it could still be beneficial to split stocks and ETFs into a few subclasses each, as in clustering approaches, and take baseline averages within these subclasses. The minor differences observed among the $RPS$ values of the top-performing teams (Makridakis, et al., 2023), however, suggest that there might not be much room for improvement in this direction. In the case of investment decisions, a potential avenue for improvement would be to optimize long-term and short-term times individually for computing each Stock trend rather than uniformly selecting 120 and 40 days, respectively. We have included only information within the M6 competition universe. Therefore, a limiting aspect of the approach is the lack of available information, such as historical data for other asset prices, financial indicators, and other signals that could impact asset prices.

## Acknowledgments

J.M.G.V. acknowledges support from Ministerio de Ciencia e Innovación under grant PID2021-128850NB-I00 (MCI/AEI/FEDER, UE) and the Basque Government under grant IT1745-22.

## Code availability

The code needed to reproduce the results is available as a python notebook at https://github.com/jmgvilar/M6competition

## Figures

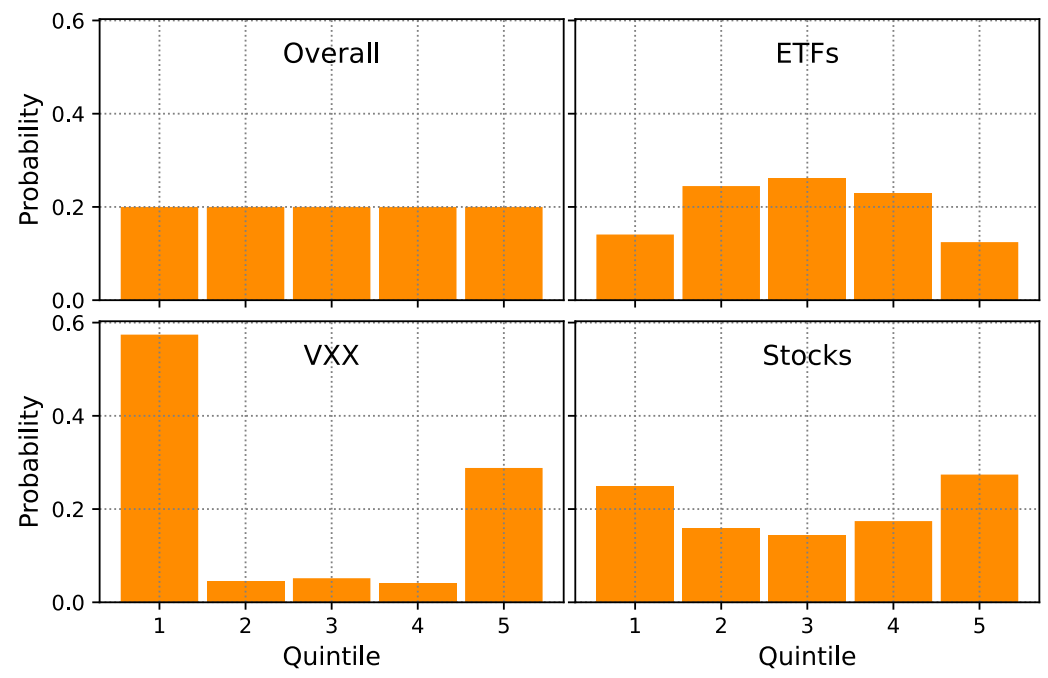

**Fig. 1.** Probability estimates of observing the returns of an asset type in each quintile. The histograms were generated from returns for the year 2022 for all the assets (top left), only ETFs (top right), only VXX ETF (bottom left), and only stocks (bottom right).

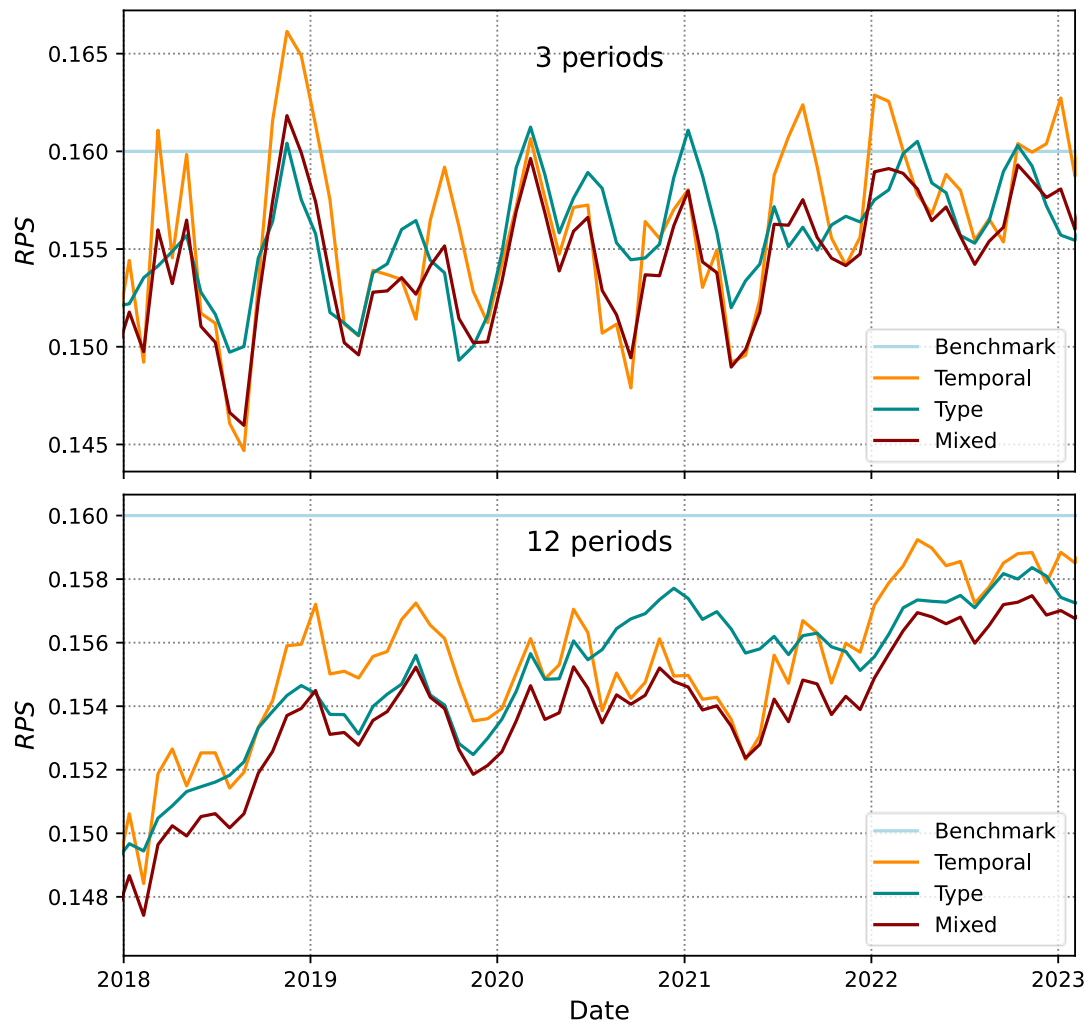

**Fig. 2.** Ranked probability scores ($RPS$) for different average procedures and moving time intervals. RPS values are computed for a moving time interval of 12 weeks (3 periods, top) and 48 weeks (12 periods, bottom) for different procedures used to estimate the probability of the ranks for each asset: overall probability estimate (benchmark), time average for each asset (temporal), separate estimates for stocks and ETFs (type), and combination of type and temporal averages (mixed).

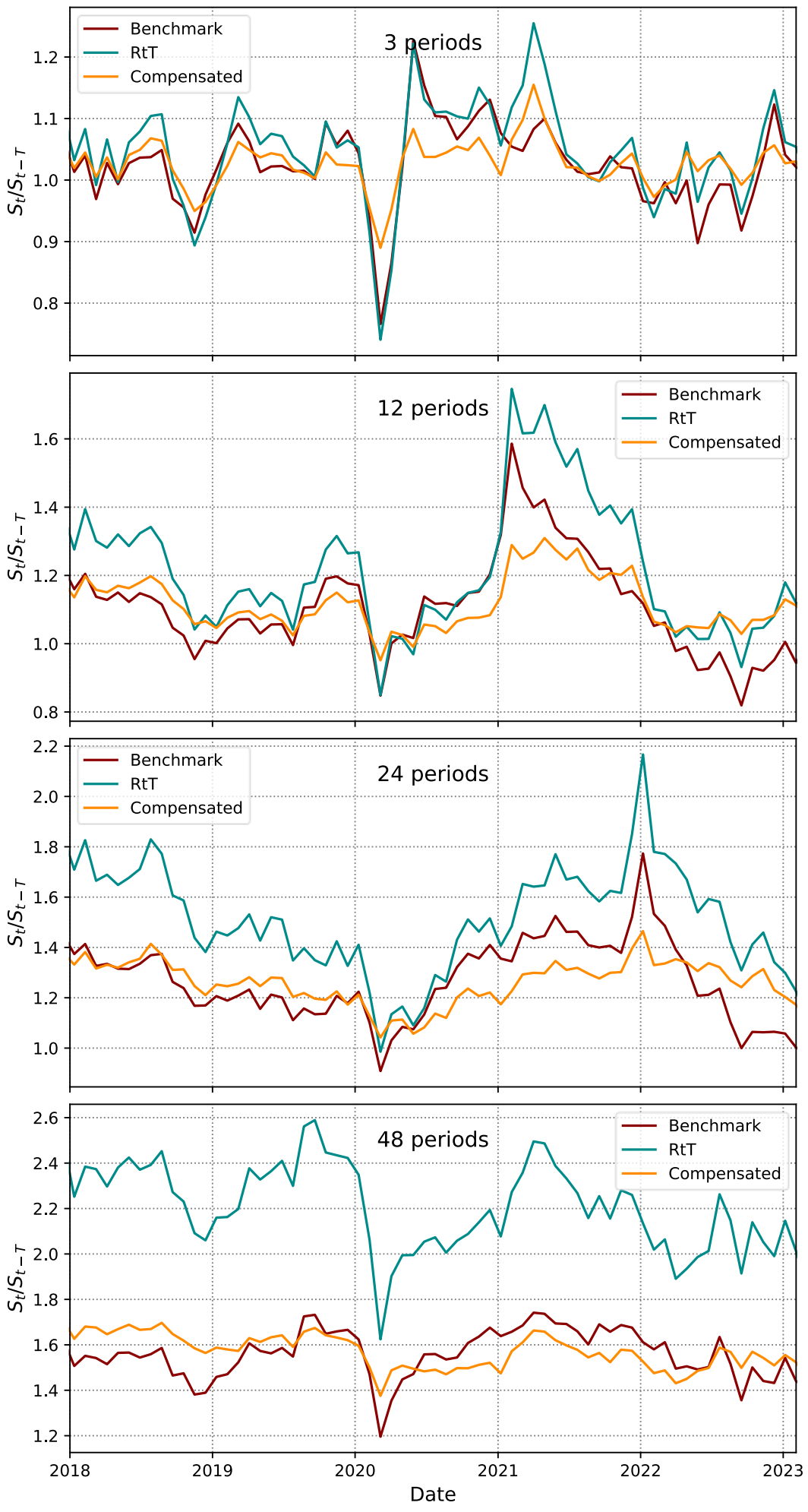

**Fig. 3.** Returns over 12 weeks (3 periods, top), 48 weeks (12 periods, second from top), 96 weeks (24 periods, second from bottom), and 192 weeks (48 periods, bottom) for the benchmark, return to the trend (RtT), and compensated portfolios.

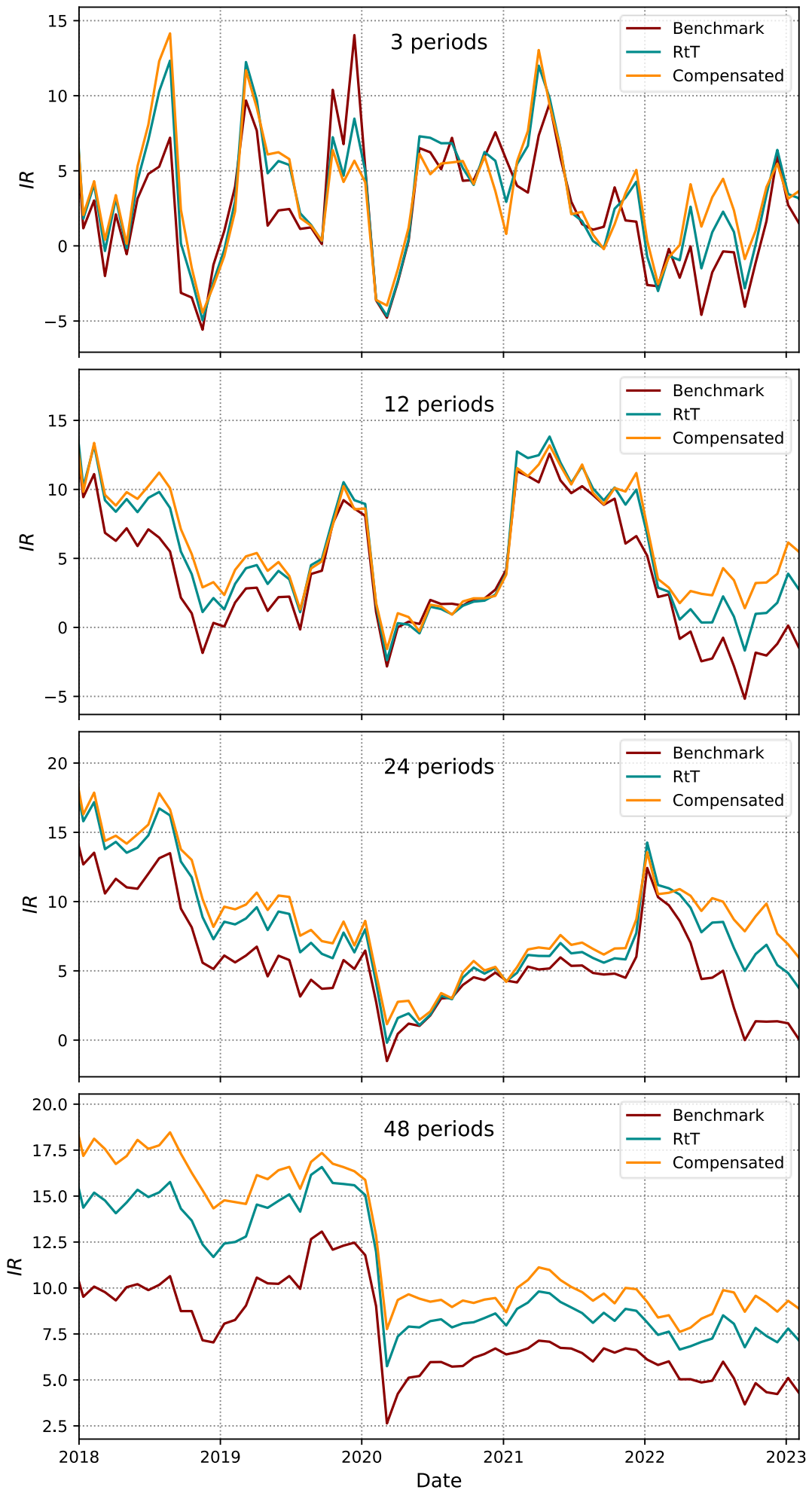

**Fig. 4.** Information ratio ($IR$) over 12 weeks (3 periods, top), 48 weeks (12 periods, second from top), 96 weeks (24 periods, second from bottom), and 192 weeks (48 periods, bottom) for the benchmark, return to the trend (RrT), and compensated portfolios.

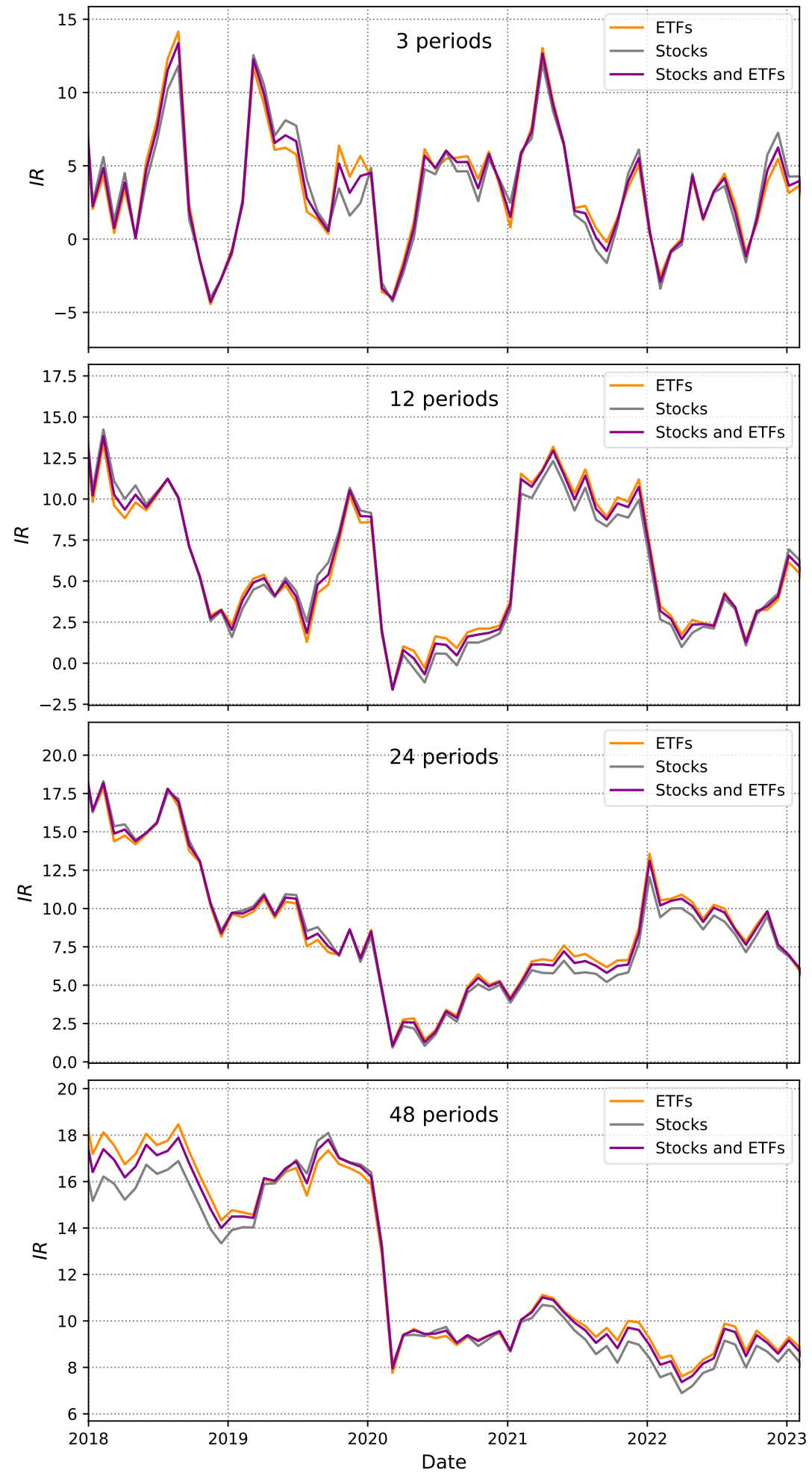

**Fig. 5.** Information ratio ($IR$) over 12 weeks (3 periods, top), 48 weeks (12 periods, second from top), 96 weeks (24 periods, second from bottom), and 192 weeks (48 periods, bottom) for compensated portfolios with just ETFs, as in Fig. 4, just stocks, and both stocks and ETFs.